\documentstyle[prl,aps,twocolumn,floats]{revtex}
\begin{document}
%\draft
\title
{A Model for  Luminous and Long Duration Cosmic Gamma Ray Bursts}

\author{Abhas Mitra}
\address{Theoretical Physics Division, Bhabha Atomic Research Center,\\
Mumbai-400085, India\\ E-mail: amitra@apsara.barc.ernet.in}

%\date{\today}

\maketitle

\begin{abstract}
We present here a simple and generic model for the  luminous ($Q_\gamma
\ge 10^{51}$ erg) and long duration ($t_\gamma \sim 10$s) Gamma Ray Bursts
(GRBs) based on the {\em fundamental fact} that the General Theory of
Relativity (GTR) suggests the existence of Ultra Compact Objects (UCOs) having
surface gravitational red-shift $z_s \le 0.615$ even when {\em most stringent}
constraint is imposed on the equation of state. This simple model may
explain the genesis of an electromagnetic fireball (FB) of energy as high
as $Q_{FB} \sim 5\times 10^{53}$ erg and an initial bulk Lorentz  Factor
as high as $
\eta \sim 10^3$.

\end{abstract}

\vskip 0.5cm
It is now clear that a large number of Gamma Ray Bursts (GRB) involve
emission of $\gamma$-rays as large as  $Q_\gamma \sim 10^{52}
-10^{53}$ erg under condition of isotropy\cite{1}.  Afterglow observations
of  GRB970228, 970508 and 980703 show  that they  indeed have quasi spherical
morphology\cite{1,2}. In fact, if GRB9901023 were also isotropic, one would infer a
value of $Q_\gamma \approx 3.4 \times 10^{54}$ erg\cite{1}! However, in
this paper we shall not consider the unique case of GRB9901023, which
might be anisotropic\cite{1},
and focus
attention on the (other) most luminous events recorded so far. We explain
them below as 
 events related to the formation of UCOs
whose {\em existence is suggested by GTR} irrespective of the details of
the EOS of the collapsing matter.  Since this is a spherical
model, unlike the irregular non-spherical models, the liberated energy
{\em will be in the form of photons and neutrinos alone}, and, not in
gravitational radiation.

As was first shown by  Schwarzschild in 1916\cite{3}, 
GTR yields an absolute upper limit on the value of the surface
gravitational redshift of a {\em static} relativistic spherical star
irrespective of the details of the Equation of State (EOS):
\begin{equation}
z_s =   \left (1 - {2G M(R)\over Rc^2}\right)^{-1/2} -1 \le z_c=2
\end{equation}
Here the subscript ``s'' refers to the respective ``surface'' values, $R$
is the invariant circumference radius, $c$ is the speed of light, and $M$
is the gravitational mass inclosed within $R=R$
\begin{equation}
M(R) = \int_0^R \rho dV =\int_0^R dM
\end{equation}
where $\rho$ is the total mass-energy density, $dV=4\pi R^2 dR$ is coordinate volume
element, and the symbol $dM$ is self-explanatory. 
Schwarzschild   obtained this limit for homogeneous
stars by {\em demanding that the central pressure of the star does not blow up}.
This result is actually  valid even for
non-homogeneous stars (see
 pp.333 of Weinberg)\cite{3} and
 is obtained when the EOS  is
allowed to have a causality violating sound speed $c_s =(dp/d\rho)^{1/2}
>c$. When the EOS is constrained to obey
causality, it follows from Eq.(9.5.19) of
Shapiro \& Teukolsky (pp.261)\cite{4}, that one would have a tighter
 limit on $z_c=
 1.22$. If one constrains the EOS further so as to have $c_s
\le c/\sqrt{3}$, it follows\cite{5} that one has 
an even tighter bound on $z_c = 0.615$. To present a
{\em realistic model}, in the following we shall work with the tightest
GTR and EOS bound on $z_s=z_c=0.615$. It may be also noted that this limit on $z_s$ is
{\em independent on the precise value} of $M$ itself. Thus, this limit may be applicable
to both stellar mass compact objects like Neutron Stars (NSs) or even
 supermassive stars, and, hence, it was debated 
 after the discovery of quasars, whether
their redshifts
were of gravitational origin\cite{3}.
 Note that, the {\em presumed} canonical NS has a value of $M
\sim 1 M_\odot$ and $R\sim 10$km with $2GM/ Rc^2 \sim 0.26$
and $ z_s \sim 0.16$.
However, actually,  many  existing EOSs  easily allow a value of $M=
(2-3) M_\odot$ and $R \approx 7$Km (for degenerate matter, there is inverse
relationship between $M$ and $R$)\cite{4}, and this may result in a value of
$2GM/ Rc^2 \approx 0.63$ or $z_s \approx 1.0$.
Thus, we can very well have an UCO, in lieu of a so-called NS. And our UCO
is {\em nothing but a NS} having a compactness, though, higher than the canonical
(i.e, {\em assumed}) value, very much allowed not only GTR but also by all
 existing EOSs. So, all that GTR tells here is that for a non-trivial
($p\neq 0$) EOS, 
{\em gravitational collapse process may end with static objects} having
$z\le z_c$. Beyond,
$z >z_c$, there would be no stable static configuration. However, light
can escape from the collapsing body and the collapse is, in principle,
reversible, until one crosses a deadline of $2GM/Rc^2 =1$ or $z=\infty$,
when the collapse becomes irreversible. However, because of inaccuracies,
the present day numerical computations have a blurred vision about $z_c$,
and further, erroneously, they conclude that the collapse becomes
irreversible immediately after $z=z_c\approx 0.16$.

Recall that the 10 GTR collapse equations form a set of highly complicated
coupled non-linear partial differential equations. Even, numerically,
 they can be solved to obtain a {\em unqiue} result only for the
homogeneous dust ($p=0$).  Other cases might
also be solved upto a certain extent but there could be hundreds of
solutions depending on a number of explicit or implicit assumptions one
makes, like self-similarity, adiabaticity, polytropic EOS, the variation
of the polytropic index, nature of opacities, radiation transport properties
{\it etc, etc}. Equally important is the question of the initial
conditions one explicitly or implicity assumes. And then depending on the
expectations, one might get the desired result.  
The genesis of a high $z_s$ object  would
be
 marked by emission of energy flux $Q \sim
Mc^2$ and then it becomes practically impossible to handle the most complex
coupled energy transport problem in a precise manner. 
By definition, in such cases, one  requires to work in the
strong gravity limit where most of the inherent assumptions break down.
For stellar mass objects, although, the high density {\em cold EOS} is
known with relatively  more certainty, our knowledge about finite
temperature EOS of nuclear matter at arbitrary high $T$ is, at present, at
is infancy. 
Also note that for numerical
computations,  for a total accumulated uncertainty of few percentage,
arising from either present theoretical inputs and intrinsic
simplifications, a potential result like $2 GM/R c^2 \approx 8/9$
(corresponding to $z_s \approx 2$, with {\em finite} gravitational
acceleration) may precipitate to a ``$2GM/R c^2 \approx 1$'' ($z \approx
\infty$ with {\em infinite gravitational acceleration}) signalling the
apparent formation of an early ``event horizon'' or a ``trapped
surface''.
Even if we consider the infinitely simpler problem of collapse of an {\em
inhomogeneous} dust, there could be varied numerical results, and, in
particular, there is a raging debate whether such collapse gives rise to a
Black Hole or a ``{\em naked singularity}''\cite{6}.
Such gross uncertainties may, at present, obfuscate the signals of
formation of more compact NSs.
In any case, as dicussed before, {\em  both the nuclear EOSs and GTR
actually suggest the
existence of} more compact NSs.

The self-gravitational energy of a static relativistic star is given by\cite{3}
\begin{equation}
E_g= \int \rho c^2 dV \left\{1- \left[1- {2GM(R)\over Rc^2}\right]^{-1/2}\right\}
\end{equation}
 Then recalling the definition of $z$ from Eq.(1), we may write
 \begin{equation}
E_g= -\int z(R)c^2 dM \approx -\alpha z_s Mc^2 \sim -z_s Mc^2
\end{equation}
where $\alpha \sim 1$ is a model dependent parameter. The binding energy,
i.e., the energy liberated in the formation of the eventually cold stellar
mass compact object, is given by virial theorem to be
$E_B \approx (1/ 2) \mid E_g \mid$.
Most of this binding energy is expected to be radiated in the form of $\nu-\bar\nu$ during
the final stages of formation of the UCO:
\begin{equation}
Q_\nu \approx E_B \approx {z_s Mc^2 \over 2}
\end{equation}
So, given the most restricted limit $z_c=0.615$ the maximum value of
$Q_\nu
\approx 0.6 M_\odot c^2 M_2 \approx 1.2\times 10^{54} M_2 $ erg where of $M= M_2
 2 M_\odot$.  This is in
agreement with our similar previous crude estimate\cite{7}.
 The value of
$Q_\nu$ measured near the compact object will be higher by a factor
$(1+z_s)$:
$Q_\nu' =  z_s (1+z_s) M/2$ (now we set $c=1$).
For the NS-formation case, the neutrinos diffuse out of the hot core in a
time $t_\nu < 10$s and we may expect a somewhat longer time scale for the
diffusion of neutrinos from the nascent hot UCO. However, here note that,
the rather long value of $t_\nu <10$ s occurs because of {\em coherent
scattering} of neutrinos by the heavy (Fe) nuclei\cite{4}. If the Fe-nucleons are
already {\em partially dissociated} by an immediately preceding heating, the
rise in the value of $t_\nu$ for the UCO formation need not be much larger.
And the locally
measured duration of the burst would be $t_\nu'=(1+z_s)^{-1} t_\nu$.
Therefore, the mean
(local) $\nu -\bar \nu$ luminosity    will be
\begin{eqnarray}
L_\nu' &=& {Q_\nu'\over t_\nu'}= {z_s (1+z_s)^2 M \over 2 t_\nu} \nonumber\\
& &{}\approx 2\times 10^{53} z_s (1+z_s)^2
M_2~ t_{10}^{-1} ~erg/s
\end{eqnarray}
where  $t_\nu =t_{10}  10 s$. 
It may be noted that this value  of $L_\nu'$ is 
well below the corresponding $\nu$-Eddington luminosity\cite{4,8}.
The luminosity in each flavour will be
 $L_i' =(1/3) L_\nu'$.
By assuming the radius of the neutrinosphere to be $R_\nu \approx R$, 
the value of effective local neutrino temperature $T'$ (assumed to be same
for all the flavors), is obtained from the condition
\begin{equation}
L_\nu' = {21\over 8} 4\pi R^2 \sigma T'^4
\end{equation}
where $\sigma$ is the Stephan-Boltzman constant. Therefore, we have,
\begin{eqnarray}
T' &=& \left( {2  z_s (1+z_s)^2 M c^2\over 21 \pi \sigma R^2 t_\nu}\right)^{1/4}\nonumber\\
& &{}\sim 13.3 MeV  z_s^{0.25} (1+z_s)^{0.5}~ M_2^{0.25} ~R_6^{-0.5} ~t_{10}^{-0.25}
\end{eqnarray}
where $R= R_6 10^6$.
For a Fermi-Dirac distribution,
the mean (local) energy of the neutrinos is $E_\nu' \approx
3.15  T' \sim 48$ MeV (for $z_s=.6$).
The various neutrinos will collide with their respective antiparticles to
produce electromagnetic pairs by the $\nu +\bar\nu \rightarrow e^+ + e^-$
process. The rate of energy generation by pair production per unit volume
per unit time,  at a distance $r$ from the center of the star, is
given by\cite{9}:
\begin{equation}
{\dot q}_\pm (r) = \sum_i {K_{\nu i} G_F^2 E_\nu' L_i'^2(r) \over 12
\pi^2 c R_\nu^4} \varphi(r)
\end{equation}
Here, $L_i'(r)\sim r^{-2}$ is the $\nu$-luminosity of a given flavour above the
$\nu$-sphere,   $G_F^2= 5.29 \times 10^{-44}~cm^2~ MeV^{-2}$ is the universal Fermi weak coupling
constant squared,  $K_{\nu i} =2.34$ for electron neutrinos and has a value of
0.503 for muon and tau neutrinos. Here the geometrical factor $\varphi(r)$ is 
\begin{equation}
\varphi(r) = (1-x)^4 (x^2 +4x +5); \qquad x= [1- (R_\nu/r)^2]^{1/2}
\end{equation}
Now, considering all the 3 flavours, a simple numerical integration yields the
 local value of pair luminosity produced above the neutrinosphere :
\begin{eqnarray}
L_\pm' &=&  \int_R^\infty {\dot q}_\pm 4 \pi r^2 dr \approx 
\sum_i { K_{\nu i} G_F^2  E_{\nu,l} L_{\nu,l}^2 \over 27
\pi c R_\nu} \approx  7 \times 10^{51} \nonumber\\
& &  ~z_s^{2.25}~(1+z_s)^{4.5}~ M_2^{2.5}~t_{10}^{-2.5}~R_6^{-2} erg/s
\end{eqnarray}
This estimate is obtained by assuming rectilinear propagation of neutrinos
near the UCO. Actually,  in the strong gravitational field near the UCO
surface the {\em neutrino orbits will be curved} with
significant higher effective interaction cross-section. Since, most of the
interactions take place near the $\nu$-sphere, for a modest range of
$z_s$, we may tentatively
try to incoroprate this nonlinear effect by inserting a $(1+z_s)^2$ factor in the
above expression.
On the other hand, the value of this electromagnetic luminosity measured
by a distant observer will be smaller by a factor of $(1+z_s)^{-2}$, so
that eventually, $L_\pm =L'_\pm$ of Eq.(11).
And the total energy of the electromagnetic FB at $\infty$ is
\begin{eqnarray}
Q_{FB}  &=& t_\nu ~L_{\pm}= 7 \times 10^{52} \nonumber\\
&=&   ~z_s^{2.25}~(1+z_s)^{4.5}~ M_2^{2.5} ~t_{10}^{-1.5}~R_6^{-2}~erg/s
\end{eqnarray}
Thus, the {\em efficiency} for conversion of $Q_\nu$ into $Q_{FB}$ is
\begin{eqnarray}
\epsilon_\pm &=& {Q_{FB} \over Q_\nu} \nonumber\\
& &\approx 3.3\% 
~ z_s^{1.25} (1+z_s)^{4.5} M_2^{1.5}
~t_{10}^{-1.5} ~ R_6^{-2}
\end{eqnarray}
In particular, for $z_s=0.615$, $M_2 =1$, $R_6 =1$ and $t_{10} =1$, we
obtain a large $\epsilon_\pm \approx 15.5\%$, and it may be reminded here
that the value of $\epsilon_\pm$ should saturate to a limiting value of
$\sim 40\%$, corresponding to a local statistical equilibrium between the
3 flavours of $\nu, \bar \nu$ and $e^+, e^-$.
This highest value of efficiency may be attained, for instance,
for $R =7$km and $M=2.5 M_\odot$.
Correspondingly, we obtain a {\em highest value} of
$Q_{FB} \approx 4.8 \times 10^{53} M_2$ erg in this model.
And thus we may  explain the energy budget of GRB971214,
$Q_\gamma \approx 3\times 10^{53}$ erg\cite{1}
 {\em without  overstretching any theory or
making any unusual assumption or invoking any unconfirmed exotic physics}
(like  ``strange stars').

Now, we shall address the question of baryonic pollution:~ $\eta= Q_{FB} /\Delta M > 10^2$.
 In general all
 models involving collision and full/partial disruption of compact
object(s) will spew out thick and massive debris ($few~ M_\odot > M_*
> 0.1 M_\odot$). Part of this debris is likely to settle into a torus
and an uncertain small fraction ($\Delta M$) may hang around the system
and get accreted on a long time scale or may even be unbounded. It is
practically, impossible to simulate the latter fraction dynamically even
in a Newtonian theory. On the other hand, {\em spherical implosion models
are completely free from the presence of such unaccountable and intractable
 thick
collisional debris}. However, in a normal SN event (assumed to be basically spherical
implosion), the ejection of baryonic mass $\sim 0.1 M_{\odot}$ occurs probably
because of shock mediated hydrodynamic process. Since, by definition, the
system is {\em gravitationally bound}, any normal hydrodynamic attempt of mass
ejection can not be much successful in a spherical model. But the shock generates additional entropy
and heat in its vicinity and might be able to effect the mass ejection. Yet, the
shock is constantly depleted of energy and gets stalled 
because of $\nu$-losses,  and
disintegration of heavy nuclei\cite{8}. Probably, the shock might be
rejuvenated by the ``shock reheating mechanism''\cite{8}. The energy
transfer between neutrinos and matter behind the shock is mediated
primarily by the charged current reactions $\nu_e + n \rightarrow
p+e^-$ and $\bar\nu_e +p \rightarrow n+ e^+$. When these reactions
proceed to the right, the matter heats up, and conversely, the matter
cools. To have a successful and sufficient net heating is a critical
phenomenon, and present day (realististic) SN codes are unable to find the
shock mediated mass-ejection (explosion) even in a relatively {\em weak} nascent-NS
gravitational field\cite{10}.  It is not surprising then that the same numerical
calculations, at present, do not find existence of UCO,   whose study
involves strong gravity,  finite temperature EOS and complex physics. 
Probably, only for a
narrow range of initial conditions and {\em modestly deep} gravitational
potential well this mechanism of shock ejection is successful. Thus the real issue is not
how to explain the non-ejection of mass by direct hydrodynamic processes
by defying the  extremely deep relativistic potential well. On the
contrary, the meaningful question is how, for a weak Newtonian potential
well, for certain range of initial conditions, there could be successful
hydrodynamic mass ejection.  Note that an UCO with a modest value of $z_s
\sim 0.5$ has a potential well which is $\sim 300 \%$ deeper than the one
associated with a canonical NS, $z_s \sim 0.16$.
Again the basic reason  that a {\em critical phenomenon} like shock heated mass
ejection might be successful for the SN case is that as one moves from a
 relativistic potential well (high $z_s$) to a Newtonian well ($z_s \le
0.2$) is that
 while the local temperature due to $\nu$-hating may decrease slowly $T'
\sim z_s^{0.25}$ and the $\nu$-matter interaction cross-section
$\sigma_{\nu,m} \sim T'^2\sim z_s^{0.5} $, the depth of the potential well
drops rapidly $\propto z_s$. Even then, it is far from clear how the
hydrodynamic mass ejection can really occur. In fact there are ideas that
departure from spherical symmetry induced by rotation and magnetic field
might be important in effecting the SN mass-ejection\cite{4}.

On the other hand, there is a genuine possibility,  that all models of
cosmological 
GRBs, irrespective of whether they  explicitly invoke the $\nu +\bar\nu \rightarrow
e^+ +e^-$ process or not, 
should involve strong direct electromagnetic or $\nu-heated$ mass loss.
Even if an unusual pulsar is assumed to emit $\sim 10^{52}$ erg/s rather
than $10^{38-40}$ erg/s,
 {\em the superstrong return current
impinging back on the pulsar} may drive a catastrophic
wind, a possibility {\em overlooked} so far.
On the other hand,
for the thin outermost layers of the object (UCO or an
hot accretion torus) emitting the neutrinos, well above the $\nu$-sphere, the
$\nu$- flux $S_\nu$ may induce a super-Eddington photon flux $S_{ph}$\cite{11}.

Also,  though, for a torus with uncertain dynamically changing geometry,
 it is difficult to make any  semi-analytical or numerical 
 estimate of such a process, in general this effect is expected to be much
more pronounced because its gravitational {\em self-binding} (
 $z_s$) is much weaker than
that for a spherical UCO surface. 
 And even if a
steady state model calculation yields a high value of $\eta$, the eventual
 value of $\eta$ might be very low if the jet is intercepted by this debris.
  In fact, (by ignoring such unmanagable real life uncertainties and difficulties),
detailed Newtonian and crude post Newtonian  calculations for the NS-NS collision case have been
presented by several authors\cite{12}; and the conclusion is that, it is difficult
to understand a value of $\eta$ higher than few.

For non-spherical configurations, it is difficult to ensure that a small fraction of the
accreted matter itself is not contaminating the FB. And the estimate of
$\Delta M$ may be
made with much larger confidence  only for a spherical model, where
by definition, {\em the entire matter, in general, is moving inwardly}. 
Here, a certain fraction of the matter lying above the neutrinosphere may
be ejected out by the $\nu$-heating
and it may be possible to crudely estimate the
 baryonic mass lying above
the $\nu$-sphere independent of the details of the problem. The mean cross section for
$\nu_e$-matter interaction is approximately given as\cite{8}
\begin{equation}
\sigma_{\nu,m} \approx 9 \times 10^{-44} (E_\nu'/1 MeV)^2 ~cm^2
\end{equation}
so that, given our range of $E_\nu' \sim 40-50$ MeV, the value of
$\sigma_{\nu, m}  < 10^{-40}$ cm$^2$.  Since the $\nu$-optical
thickness of the layer above the $\nu$-sphere is $\sim 2/3$, the surface
density of this layer $\delta \sim 2m_p/(3
\sigma_{\nu,m}) \sim 10^{16}$ g/cm$^2$, where $m_p$
is the proton rest mass. Therefore, the mass of the matter above the
$\nu$-sphere is $\Delta M \sim 4 \pi R_\nu^2 \delta \sim 10^{29}~g \sim
10^{-4} M_{\odot} R_6^{2}$.
 Probably, the most detailed work on this problem of $\nu$-driven mass
ejection from a hot nascent 
 NS is due to
Duncan, Shapiro \& Wasserman\cite{11}; and the {\em Table 5} of it shows that for
$R\approx 10^6$ cm, $M=2 M_\odot$, we have $\Delta M \approx 10^{-4}
M_\odot$, if $T'= 20 MeV$. On the other hand, for $T' =30 MeV$, one has,
$\Delta M \approx 7 \times 10^{-4} M_{\odot}$. This estimates were made
in the framework of Newtonian gravity, and a GTR calculation, if possible,
would certainly yield,  lower values of $\Delta M$. Even considering
 these Newtonian values of $\Delta M$, we find that the value of $\eta$
could be easily lie between $10^3 > \eta > 10^2$.
Now the occurrence of  luminous and long GRBs can be understood by using the
existing ideas\cite{13}. 

Previously, accretion induced collapse of a White
Dwarf to a NS was suggested as a model for GRBs\cite{9}. The difficulty of this
model was that (i) 
 for a canonical NS with $M =1 M_\odot$ and $z_s\approx
0.16$,  Eq.(12) yields
a very low value of $Q_{FB} \approx 4\times 10^{50}$ erg, (ii) further, the
value of $\eta$ is seen to be too low, and (iii) in the corresponding {\em
weak}
gravitational well, it can not be ensured that supernova shock is not launched.

And it is probable that less luminous and short
GRBs may occur by various other scenarios too including the so-called ``collapsar''
or ``hypernova'' type models\cite{14}.
which, may explain the origin of a value of
$Q_\gamma \sim 10^{49} -10^{51}$ erg for a duration of $t_\gamma < 1
$s (though in extreme cases they go beyond this range).

\end{document}